\documentclass[%
 preprint,
 singleocolumn,
 amsmath,amssymb,
 aps,
 pre,
showpacs]{revtex4-1}

\usepackage{graphicx}
\usepackage{dcolumn}
\usepackage{bm}
\usepackage{subfigure}
\usepackage{float}
\usepackage{color}
\usepackage{array}
\usepackage{float}
\usepackage{multirow}

\begin{document}


\title{Hidden jerk in universal creep and aftershocks}

\author{Vikash Pandey}

\email{vikash4exploring@gmail.com}
\affiliation{%
School of Interwoven Arts and Sciences, Krea University, Sri City 517646, India
}%

\date{\today}

\begin{abstract}
Most materials exhibit creep memory under the action of a constant load. The memory behavior is governed by Andrade's creep law, which also has an inherent connection with the Omori-Utsu law of earthquake aftershocks. Both empirical laws lack a deterministic interpretation. Coincidentally, the Andrade law is similar to the time-varying part of the creep compliance of the fractional dashpot in anomalous viscoelastic modeling. Consequently, fractional
derivatives are invoked, but since they lack a physical interpretation, the physical parameters of the two laws extracted from curve fit lack confidence. In this Letter, we establish an analogous linear physical mechanism that underlies both laws and relates its parameters with the material’s macroscopic properties. Surprisingly, the explanation does not require the property of viscosity. Instead, it necessitates the existence of a rheological property that relates strain with the first order time derivative of stress, which involves jerk. Further, we justify the constant quality factor model of acoustic attenuation in complex media. The obtained results are validated in light of the established observations.




\vspace{100 pt}
\texttt{The peer-reviewed version of this manuscript is published as a, \textit{Letter}, in the journal, Physical Review E, Vol.~107, No.~2, Pages~L022602, 2023.} \\
DOI: 
\href{https://doi.org/10.1103/PhysRevE.107.L022602}{https://doi.org/10.1103/PhysRevE.107.L022602}. 

\texttt{The published version of the manuscript is available online
at 
\\\href{https://link.aps.org/doi/10.1103/PhysRevE.107.L022602}{https://link.aps.org/doi/10.1103/PhysRevE.107.L022602}.}

\textcolor{blue}{\textbf{This document is an e-print which may differ
in, e.g. pagination, referencing styles, figure sizes, and typographic
details.}}
\end{abstract}

\maketitle


Most materials continuously deform under constant exposure to load and
eventually fail. This mechanical failure is at the core of natural
calamities, e.g., avalanches, landslides, and earthquakes. Early
forecasting of a possible failure could mitigate the catastrophic
consequences that have their origin in natural causes and engineered ones, for example, the collapse of bridges.
Although the deformation mechanism is complex, simple explanations using power laws with few parameters are preferred. They have been used to describe the creep in materials \cite{Main2000,Dysthe2002,Tsai2016,Benzi2021,Dijksman2022,Head2022}
and the rate of earthquake aftershocks \cite{Baro2013,Ribeiro2015,Davidsen2017,Lherminier2019}. 

The century-old Andrade creep law \cite{Andrade1910,Nicolas2018} describes a material's primary creep response, $\varepsilon\left(t\right)$, due to constant
stress as
\begin{equation}
\varepsilon\left(t\right)=\varepsilon_{0}J\left(t\right), \text{ where } J\left(t\right)=\left(\frac{t}{\tau}\right)^{\alpha},\label{eq:Andrade_law}
\end{equation}
$\varepsilon_{0}$ is the stress-dependent initial strain at a time, $t=0$, and $\tau$ is the characteristic retardation
time constant of the material. The exponent, $\alpha$, is around $1/3$ for soft metals \cite{Miguel2002}, and it lies between zero and one for most heterogeneous materials, including amorphous solids \cite{Nicolas2018}
and biological materials \cite{Leocmach2014,Kobayashi2017,Bonfanti2020}. Since the rupture time is proportional to the duration
of the primary creep \cite{Nechad2005}, it becomes imperative to
investigate the mechanism underlying the Andrade law. The first
few attempts to understand the creep law were motivated by nonlinear models, for example, Schapery's
stress-strain constitutive equation, but
it is plagued by four unknown material parameters with questionable
reproducibility \cite{Gamby1987}. Another model used a parallel combination of a spring and a nonlinear Eyring dashpot to numerically predict the particular
case, $\alpha=1$ \cite{Nechad2005}. The mathematical complexities
that arise due to the nonlinearity have not proved beneficial.
Lately, fiber bundle models \cite{Roy2018,Castellanos2019} have been examined
in the stochastic framework, ignoring that \textit{plasticlike}
primary deformation is deterministic. Since a direct relation between a material's properties and the parameters $\tau$, and $\alpha$, is not yet established, their values are extracted from the curve fit of experimental data with the theoretically predicted curves from Eq.~$\left(\ref{eq:Andrade_law}\right)$  \cite{Sassinek2014}. A computationally
intensive approach could probably simulate the power-law creep
in the framework of nonlocal elasticity \cite{Jagla2014,Pandey2016}, but an interpretation of the parameters is not guaranteed.

The discrete counterpart of the Andrade law is the Omori law, and
it expresses the rate of occurrence, $R$, of earthquake aftershocks
as a function of time since the main shock  \cite{Omori1894}. Later, the law was  generalized
as the Omori-Utsu (OU) aftershock law \cite{Utsu1961,Hergarten2002}: 
\begin{equation}
R\left(t\right)=\frac{\chi}{\left(c+t\right)^{p}}.\label{eq:Omori_law}
\end{equation}
The OU law shares an inherent connection \cite{Castellanos2019} with the Andrade law as
$R\left(t\right)\sim dJ\left(t\right)/dt$, so the Omori exponent, $p\sim1-\alpha$, the time-constant, $c\sim\tau$, and the productivity, $\chi\sim\alpha\tau^{\left(p-1\right)}$. Although $p$ is generally
around one, values between $0.5$ and $1.6$ have also been
reported \cite{Sornette2005,Bottiglieri2010,Davidsen2011}.  Since the OU law
represents the decay activity after a significant disturbance has occurred
in a system, it has been used to model avalanches \cite{Bak2002},
universality in solar flares \cite{Arcangelis2006}, epileptic attacks
\cite{Osorio2010}, and stock market dynamics \cite{Petersen2010a}.
Probably the OU law was first obtained numerically using
the Carlson-Nager nonlinear model \cite{Carlson1989,Sakaguchi2015}, which was motivated by the stick-slip mechanism of friction \cite{Dieterich1978}. Another nonlinear model based on stress-induced corrosive damage mechanics unified the Andrade law and the OU law \cite{Main2000}.
Few explanations stem from the mathematical
result approximating a power law from the probability density
of the broad distribution of characteristic waiting times between
consecutive events \cite{Lindman2005}. However, that can only predict the
specific case, $p=1$ \cite{Baro2013,Baiesi2004}. It is not a formal proof of the law because the converse is not always true
\cite{Saichev2006}. Besides, such distributions' breadth and form are difficult to comprehend. An explanation based on self-organized
criticality lacked a deterministic
interpretation because of its
inherent statistical nature \cite{Bak2002}. Some  works connected the OU law with the specimen
geometry \cite{Tsai2016}, microscopic dynamics \cite{Davidsen2017},
thermal noise \cite{Saichev2005}, mean field theory \cite{Baro2018},
and fractal hierarchical block model \cite{Mykulyak2018}. Although
the validity of the OU law has been confirmed for six decades, its
understanding is far from complete \cite{Baro2013}. Currently, a model free from curve fit does not exist \cite{Rosti2010,Braun2014,Sentjabrskaja2015}, which is also evident from the ambiguity attached to the origin of $c$ in the OU law,
whether physical or instrumental \cite{Sakaguchi2015,Davidsen2015}.

Coincidentally, the Andrade law shares a surprising connection with the fractional dashpot, whose constitutive relation is \cite{Pandey2016a}
\begin{equation}
\sigma\left(t\right)=E\tau^{\alpha}\frac{d^{\alpha}\varepsilon\left(t\right)}{dt^{\alpha}},\label{eq:frac_consti_law}
\end{equation}
where $\sigma$ is the stress, $\varepsilon$ is the strain, $E$ is the
constant modulus of elasticity, and $\alpha\in\left(0,1\right)$ is the fractional order of the derivative. For a causal function, $f\left(t\right)$, $d^{\alpha}f\left(t\right)/dt^{\alpha}=d f\left(t\right)/dt \ast t^{-\alpha}/{\Gamma\left(1-\alpha\right)}$, where $\ast$ represents the convolution operation, and $\Gamma\left(\cdot\right)$ is the Euler Gamma function \cite{Holm2019}. The fractional dashpot plays a versatile role in anomalous viscoelastic modeling  as it interpolates between a Hookean spring and a Newtonian dashpot in the limits as $\alpha\rightarrow0$ and as $\alpha\rightarrow1$, respectively. The motivation to adopt the framework of fractional derivatives is three fold. First, we know that a convolution in the time domain translates to a product in the Laplace domain, $s$, so $\mathcal{L}\left\{ d^{\alpha}\varepsilon\left(t\right)/dt^{\alpha}\right\} =\left[s\bm\varepsilon\left(s\right)-\varepsilon_{0}\right]/s^{1-\alpha}$, where $\bm\varepsilon\left(s\right)=\mathcal{L}\left\{\varepsilon\left(t\right)\right\}$, and $\varepsilon_{0}$ is the instantaneous elastic strain at a time, $t=0$. Since creep compliance corresponds to the creep response to the input of constant stress, $\sigma_{0}$, we have $\varepsilon_{0}=\sigma_{0}/E$, and $\mathcal{L}\left\{ \sigma_{0}\right\} =\sigma_{0}/s$. Substituting the Laplace transforms of the respective terms in  Eq.~$\left(\ref{eq:frac_consti_law}\right)$, the solution in the $s$ domain is $\bm\varepsilon\left(s\right)=\varepsilon_{0}\left[1/s+1/\left(\tau^{\alpha}s^{1+\alpha}\right)\right]$. The inverse Laplace transform yields $\varepsilon\left(t\right)=\varepsilon_{0}\left[1+\left(t/\tau\right)^{\alpha}/{\Gamma\left(1+\alpha\right)}\right]$, from which the time-varying part turns out to be similar to the Andrade law. Interestingly, though Scott Blair and Reiner had proposed Eq.~$\left(\ref{eq:frac_consti_law}\right)$ to describe the Nutting law \cite{Pandey2016a}, they were probably unaware of its connection with the Andrade law. Second, the observations support $\alpha\neq1$ in Eq.~$\left(\ref{eq:Andrade_law}\right)$,
and $p\neq0$, in Eq.~$\left(\ref{eq:Omori_law}\right)$, implying that
creep and aftershocks are memory-laden non-Poissonian events \cite{Shcherbakov2005}. So, failure is not an instantaneous
process; instead, it accumulates over time. Fortunately, memory is embedded in the definition of a fractional derivative in the form of the temporal power-law kernel. Fractional derivatives are also preferred because they offer a  succinct representation
of material behavior, usually described using linear-system theory involving convolutions between integer-order
time derivatives and time-dependent coefficients \cite{Pandey2016a,Pandey2022a}.  Moreover,
the Fourier transform property, $\mathcal{F}\left[d^{\alpha}f\left(t\right)/dt^{\alpha}\right]=\left(i\omega\right)^{\alpha}\boldsymbol{F}\left(\omega\right)$,
where $\boldsymbol{F}\left(\omega\right)$ is the Fourier transform of $f\left(t\right)$
in the frequency domain, $\omega$,
indicates that fractional derivatives are a natural generalization
of the Newtonian derivatives. Third, the stick-slip-induced grain-shearing mechanism was successfully understood using fractional derivatives \cite{Pandey2016b}; the mechanism is common to both material deformation \cite{Dieterich1978,Zadeh2019}
and earthquake aftershocks \cite{Davidsen2017,Dieterich1972b}. Recently, fractional derivatives were
used to study the OU law, but the lack
of an interpretation of the order, $\alpha$, undermines the confidence
in those results \cite{Baro2018,Saichev2004}.

\begin{figure}
	\begin{centering}
		\includegraphics[width=1\columnwidth]{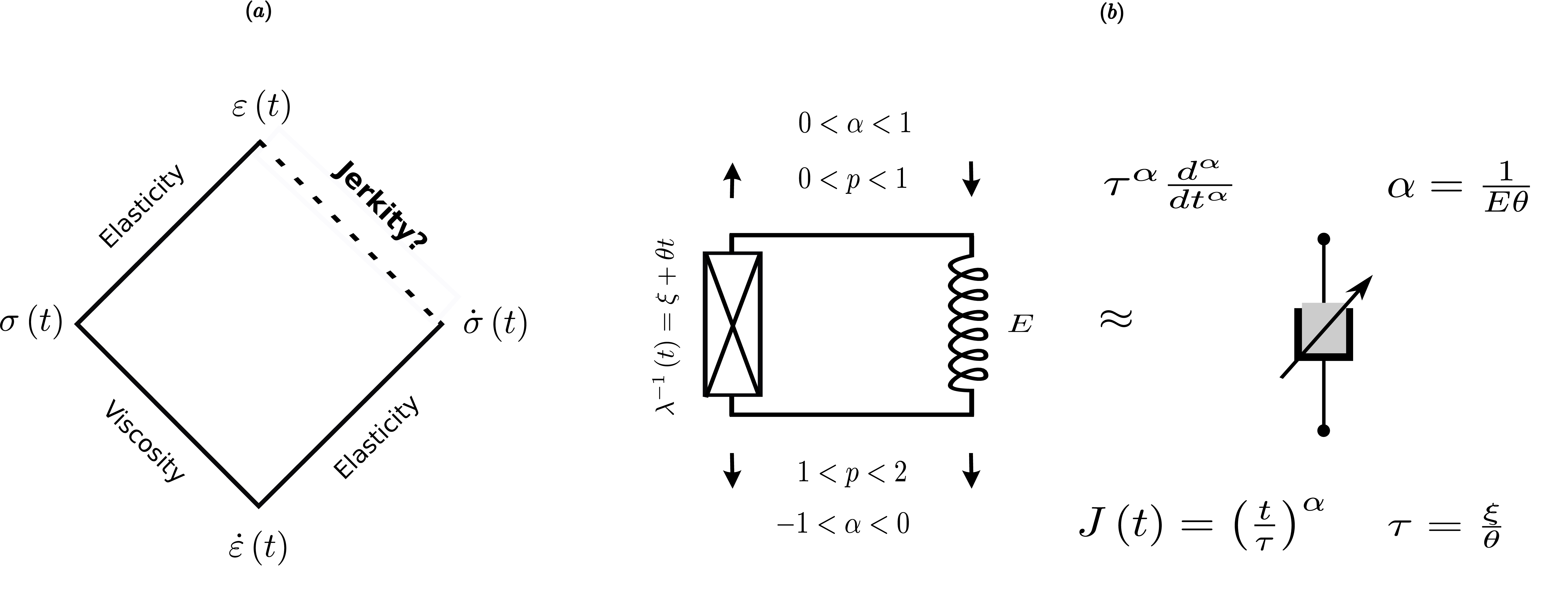}
		\par\end{centering}
	\caption{(a) Jerkity is the missing rheological property that relates $\dot{\sigma}\left(t\right)$ with $\varepsilon\left(t\right)$. (b) The creep compliance of the fractional dashpot is approximately the same as that of a parallel combination of a jerken and a spring, expressed by Eq.~$\left(\ref{eq:Andrade_law_deriv}\right)$.}
	\begin{centering}
		\label{Zerkenfig}
		\par\end{centering}
\end{figure}

As shown in Fig.~\ref{Zerkenfig}(a),  Hooke's law, $\sigma\left(t\right)=E\varepsilon\left(t\right)$, its differential form, $\dot{\sigma}\left(t\right)=E\dot{\varepsilon}\left(t\right)$, and Newton's law of
viscosity, $\sigma\left(t\right)=\eta\dot{\varepsilon}\left(t\right)$, where, $\eta$ is the coefficient of viscosity, symbolically represent the three sides of a square whose vertices are labeled as, $\sigma$, $\varepsilon$, and their respective time derivatives. The fourth side is missing because there is probably no relationship
between  $\dot{\sigma}\left(t\right)$ and $\varepsilon\left(t\right)$. In order to complete the square, we propose the property of ``jerkity," represented by the rheological element ``jerken'' with the following constitutive relation:
\begin{equation}
\frac{\dot{\sigma}_{j}\left(t\right)}{\varepsilon_{j}\left(t\right)}=\lambda,\label{eq:memring}
\end{equation}
where $\sigma_{j}$ and $\varepsilon_{j}$ are the stress and the strain in the jerken, respectively. The ``coefficient of jerkity,'' $\lambda>0$, has the units of modulus
of elasticity per unit of time. The name, \textit{jerken}, is motivated by the fact that a finite strain is possible only when $\dot{\sigma}_{j}\left(t\right)\neq0$, which corresponds to a nonzero third-order derivative of displacement, also referred to as jerk in classical mechanics. In contrast, the properties of elasticity and viscosity give a finite strain for constant positive stress. 

We justify the property of jerkity as follows. The deformation of a material is a function of its stiffness and the applied force. If the change in force is slow in time, the respective jerk is negligible. In such a case, the material experiences almost a quasistatic condition that cannot cause a wave excitation. In contrast, a time-varying force
can only cause a wave excitation, similar to creating a propagating wave pulse in a string when one of its loose ends is jerked. Andrade's observations of ``copper quakes'' also support that material deformation and aftershocks are mediated by wave diffusion \cite{Andrade1910}. This justifies the transition from a recoverable elastic deformation to an unrecoverable plastic-like deformation. Further, deformation in heterogeneous materials manifests due to stick-slip processes assuming that stresses are rarely stationary in laboratory creep tests and the earth’s crust \cite{Dieterich1978,Dieterich1972b,Rosti2009}.

Although Mott's microscopic theory intuitively described the emergence of the power law creep from  thermally-activated jerk movement of dislocations, see Fig.~2 in Ref.~\cite{Mott1953}, the jerk mechanism was probably never considered mathematically in the past theories. The constitutive relation of jerkity, Eq.~$\left(\ref{eq:memring}\right)$, yields unphysical results if not modified appropriately. For example, in the case of a positive constant, $\lambda$, the stress response of the jerken to a constant strain input is a stress that grows linearly with time, which is thermodynamically impossible. Another problem with Eq.~$\left(\ref{eq:memring}\right)$ is that in a tensile experiment, it predicts an expansion for an increasing tension, $\dot{\sigma}_{j}\left(t\right)>0$, but also an immediate contraction for a decreasing tension, $\dot{\sigma}_{j}\left(t\right)<0$. The latter implies a physically invalid negative work, verifiable by its dynamic response to the standard loading test. For the cyclic input of $\varepsilon_{j}\left(t\right)=\varepsilon_{m}\sin\left(\omega t\right)$, where $\varepsilon_{m}$ is the maximum amplitude of the strain, the corresponding stress response is $\sigma_{j}\left(t\right)=\lambda\varepsilon_{m}\left[1-\cos\left(\omega t\right)\right]/\omega\geq0$, i.e., an unphysical non-negative stress output for the input of a succession of positive and negative deformations.

The thermodynamic consistency is restored if $\lambda$ is assumed to decrease with time. A linearly increasing, $1/\lambda\left(t\right)=\xi+\theta t$, is the simplest possible choice, where $\xi>0$, 
and $\theta=d\left(1/\lambda\right)/dt\mid_{t=0}>0$, are the constant part
and the time-varying part of $1/\lambda$, respectively. The presence of $\xi$ ensures that no singularity is encountered at time, $t=0$. The time-varying nature of the material constants arises from the permanent alterations in microscopic structure due to constant exposure to mechanical, thermal, and chemical stresses \cite{Dieterich1978,Lidon2016,Kjartansson1979}. The time dependence of $\lambda\left(t\right)$ could manifest from the local stress heterogeneity that stems from the underlying fractality that is inherently present in almost all random media \cite{Lambert2015}. As the material deforms over time, its fractality and stress heterogeneity also evolve. The material deforms through a series of time-varying jerk steps.  The problem of the negative work is resolved if the jerken is connected in parallel with a spring of constant modulus of elasticity, $E$. So, an increasing $1/\lambda\left(t\right)$ and the action of a spring in parallel are the necessary conditions to make the property of jerkity physically permissible.  Thus, the model in Fig.~\ref{Zerkenfig}(b) is unique in its own right, in which the crossed-box symbol represents the jerken. Further, as the jerken is a linear  element, the  framework of superposition principles and integral transforms are applicable. 

Since the stresses due to the spring and the jerken may either oppose each other or add together, we investigate the two cases separately.
For the first case, the total stress is $\sigma_{tot}\left(t\right)=\sigma_{s}\left(t\right)-\sigma_{j}\left(t\right)$,
where $\sigma_{s}$ and $\sigma_{j}$ are the stresses in the spring
and the jerken, respectively. If the total applied stress, $\sigma_{tot}$, is a constant, then $\dot{\sigma}_{s}\left(t\right)-\dot{\sigma}_{j}\left(t\right)=0$.
As $\dot{\sigma}_{s}\left(t\right)=E\dot{\varepsilon}_{s}\left(t\right)$, 
and $\dot{\sigma}_{j}\left(t\right)=\lambda\left(t\right)\varepsilon_{j}\left(t\right)$, we have $E\dot{\varepsilon}_{s}\left(t\right)-\varepsilon_{j}\left(t\right)/\left(\xi+\theta t\right)=0$, where $\varepsilon_{s}$ is the strain in the spring.
Since the strain stays the same in the parallel branches, let $\varepsilon_{s}\left(t\right)=\varepsilon_{j}\left(t\right)=\varepsilon\left(t\right)$,
which leads to a first-order linear ordinary differential equation, $\dot{\varepsilon}\left(t\right)/\varepsilon\left(t\right)=1/\left[E\left(\xi+\theta t\right)\right]$. The integration gives $\ln\varepsilon\left(t\right)=\left[\ln\left(\xi+\theta t\right)\right]/\left(E\theta\right)+\ln C$.
At the time, $t=0$, the jerken does not experience any stress; instead, the spring takes all the applied stress, so the initial strain is $\varepsilon_{0}=\sigma_{tot}/E$. We obtain the
integration constant, $\ln C=\ln\varepsilon_{0}-\left(\ln\xi\right)/\left(E\theta\right)$,
which, when substituted back into its parent expression, gives the creep
response as:
\begin{equation}
\varepsilon\left(t\right)=\varepsilon_{0}\left(1+\theta\frac{t}{\xi}\right)^{1/\left(E\theta\right)}.\label{eq:Andrade_law_deriv}
\end{equation}
At large timescales, $\theta t/\xi\gg1$, the time-varying part of
Eq.~$\left(\ref{eq:Andrade_law_deriv}\right)$, approximates  the
Andrade law such that,
\begin{equation}
\tau=\frac{\xi}{\theta}\text{, and }\alpha=\frac{1}{E\theta}.\label{eq:frac_dash}
\end{equation}
In light of Eqs.~$\left(\ref{eq:Omori_law}\right)$, $\left(\ref{eq:Andrade_law_deriv}\right)$,
and $\left(\ref{eq:frac_dash}\right)$, we interpret the parameters of the Omori-Utsu law as:
\begin{equation}
p=1-\frac{1}{E\theta},\text{ and }\tau=c=\frac{\xi}{\theta}.\label{eq:parameters}
\end{equation}
For the second case, the total stress is $\sigma_{tot}\left(t\right)=\sigma_{s}\left(t\right)+\sigma_{j}\left(t\right)$. The
respective results are obtained as
\begin{equation}
\alpha=-\frac{1}{E\theta}\text{, and }p=1+\frac{1}{E\theta},\label{eq:omori_para}
\end{equation}
leaving $c$ unchanged. 

The derivation agrees with the prediction that the power-law creep is a superposition of two separate creep mechanisms and that $\alpha$ is independent of the applied stress \cite{Andrade1910,Andrade1962}. The creep mechanisms manifest from the interplay of the properties of elasticity and time-varying jerkity. As Andrade's law is limited to $0<\alpha<1$, it arises only in the first case, i.e., when the two mechanisms oppose each other. This is supported by the most observed value of $\alpha=1/3$ for soft metals, though such a fixed value indicates a possible interdependence between $E$  and $\theta$ for such materials. For materials with small values of $E\theta$, $\alpha$ is large, implying a fast creep. Surprisingly, although the creep law and the aftershock law were independently proposed, Andrade \cite{Andrade1910} drew parallels between geological quakes and ``copper quakes" for $\alpha=1/3$, i.e., $p=2/3$,  which suggests that the opposing nature of the two mechanisms is common to both universal creep and earthquake aftershocks. 

In contrast to the Andrade law, the OU law arises in both cases, i.e., when the two creep mechanisms due to elasticity and jerkity add up and oppose each other. In the case of the opposing nature of the two mechanisms, as mentioned in Eq.~$\left(\ref{eq:parameters}\right)$, $p=1-\alpha<1$. If the stresses due to the two mechanisms add together, then according to Eq.~$\left(\ref{eq:omori_para}\right)$,  $p=1-\alpha>1$ because $\alpha<0$. Since the thermodynamic constraints \cite{Holm2017a} limit, $0<\left|\alpha\right|<1$, the physically permissible values, $p\in\left(0,2\right)$, may be seen as a \textit{falsifiable prediction} from our model. The observed values, $0.5\leq p \leq 1.6$, are within the limits imposed by the fractional framework.  The observation that $p$ is stress independent and $\chi$ is time independent concur with Refs.~\cite{Castellanos2019} and \cite{Xu2019}, respectively.   In light of the fractional diffusion-wave equation obtained from the fractional dashpot \cite{Pandey2016b}, the diffusive wave mediates the creep in materials and the propagation of aftershock energy, which agrees with Ref.~\cite{Helmstetter2002}. Further, the magnitude of a mainshock depends on the release of the stored potential energy, which is directly proportional to the elastic constant, $E$, of the earth. Since a large value of $E$ leads to a small contribution from $\alpha$, both the expressions of $p$ from  Eqs.~$\left(\ref{eq:parameters}\right)$ and $\left(\ref{eq:omori_para}\right)$ constrain its value in the neighborhood of $p=1$ for high-magnitude mainshocks. Besides, large values of $\theta$ also favor $p=1$. The contribution from $\alpha$ for small values of $E$ is large, so a significant deviation of $p<1$ values from $p=1$ are expected for low magnitude mainshocks; for example, see Figs.~1 and 2 in Ref.~\cite{Sornette2005}. In the case of $p>1$ for intermediate to high magnitude earthquakes, as $E$ is supposed to be sufficiently large, it is expected that $p$ is close to one. However, it does not reflect strongly from the data of the California earthquake catalog mentioned in Table~\ref{Tab1}  \cite{Lennartz2008}. This anomaly can be justified if such earthquakes are characterized by small values of $\theta$ such that $E\theta$ is sufficiently small. The more significant the deviation of $p>1$ values from $p=1$ for strong earthquakes, the smaller the value of $\theta$. Moreover, since there is possibly a mutual interdependence between $E$ and $\theta$, the value of $p=1\pm 1/\left(E\theta\right)$ depends on the interplay between the creep mechanisms due to elasticity and jerkity.

\begin{table}[h]
\caption{Optimized parameters of the Omori-Utsu law for the four aftershock sequences
Parkfield, Northridge, Hector Mine, and Landers \cite{Lennartz2008}. The mainshock magnitude is in the Gutenberg-Richter scale.}
\begin{tabular}{|c|c|c|c|}
\hline 
\begin{tabular}{c}
Earthquake location\tabularnewline
(year)\tabularnewline
\end{tabular} & %
\begin{tabular}{c}
Mainshock\tabularnewline
magnitude\tabularnewline
\end{tabular} & %
\begin{tabular}{c}
$c$\tabularnewline
(days)\tabularnewline
\end{tabular} & $p$\tabularnewline
\hline 
\hline 
\begin{tabular}{c}
Parkfield \tabularnewline
(2004)\tabularnewline
\end{tabular} & $6.0$ & $0.0039$ & $1.09$\tabularnewline
\hline 
\begin{tabular}{c}
Northridge \tabularnewline
(1994)\tabularnewline
\end{tabular} & $6.7$ & $0.012$ & $1.18$\tabularnewline
\hline 
\begin{tabular}{c}
Hector Mine \tabularnewline
(1999)\tabularnewline
\end{tabular} & $7.1$ & $0.024$ & $1.21$\tabularnewline
\hline 
\begin{tabular}{c}
Landers \tabularnewline
(1992)\tabularnewline
\end{tabular} & $7.3$ & $0.08$ & $1.22$\tabularnewline
\hline 
\end{tabular}
\label{Tab1}
\end{table}

The assumption of a constant load underlying the OU law agrees with Ref.~\cite{Ribeiro2015}.  Further as $c=\xi/\theta\rightarrow0$, it implies, $\theta\rightarrow\infty$, so, $\alpha=\pm 1/\left(E\theta\right)\rightarrow0$. Hence, $p=1\pm 1/\left(E\theta\right)\rightarrow1$ gives the Omori law, i.e., $c\approx 0$ when $p\approx 1$, which agrees with Table~\ref{Tab1}. The observation \cite{Narteau2009} that stress accumulation depends on $c$ is true since $c$ is related to the rheological properties of the earth's crust through $\xi$ and $\theta$. Thus, $c$, indeed, has a physical origin, though its value could be difficult to extract in many real scenarios unless the high-frequency signals of aftershocks are carefully analyzed \cite{Peng2007}.  

We now justify the frequency independent, constant quality factor, $Q$, commonly observed in acoustic wave attenuation
in earth materials \cite{Kjartansson1979} and biological tissues \cite{Lambert2015}. The constant-$Q$ models characterized with few parameters improve seismic inversion \cite{Sun2019}. Despite the importance of those models in understanding  mantle convection currents \cite{Wesson1971}, they lacked a physical interpretation because they are inherently linked with the Andrade law, but now it is resolved as follows.  The inverse quality factor is $1/Q\left(\omega\right)=J_{2}\left(\omega\right)/J_{1}\left(\omega\right)$, where $J_{1}\left(\omega\right)-iJ_{2}\left(\omega\right)=J\left(\omega\right)=i\omega\mathcal{F}\left[J\left(t\right)\right]$ is the creep compliance function, and $i=\sqrt{-1}$ \cite{Karato2008}. The real part, $J_{1}\left(\omega\right)$, and the imaginary part, $J_{2}\left(\omega\right)$, are coupled through the Kramers-Kronig relations. Since the Kramers--Kronig relations stem from the principles of linearity and causality, constant-$Q$ models are preferred as they give an attenuation coefficient almost linearly dependent upon the frequency, which is also supported by laboratory and field measurements
\cite{Sun2019}. Using the Fourier transform property, $\mathcal{F}\left[t^{\alpha}\right]=\left(i\omega\right)^{-1-\alpha}\Gamma\left(1+\alpha\right)$, we extract, $J_{1}\left(\omega\right)\propto\omega^{-\alpha}\cos\left(\pi\alpha/2\right)$, and $J_{2}\left(\omega\right)\propto\omega^{-\alpha}\sin\left(\pi\alpha/2\right)$. So, $1/Q\left(\omega\right)\propto\tan\alpha\pi/2$, i.e., a frequency-independent constant. We impose a \textit{rigorous} test on the interpretation of $\alpha$, which is motivated by the causality principle \cite{Kjartansson1979}. The test dictates that the correct mechanism for a constant-$Q$ must predict $1/Q=0$ at zero frequency \cite{Baan2002}. The zero frequency implies $t/\tau\rightarrow\infty$, which occurs if $\tau=\xi/\theta\rightarrow0$, implying $\theta\rightarrow\infty$. So, $\alpha=1/\left(E\theta\right)\rightarrow0$, and $1/Q\propto\tan\alpha\pi/2\rightarrow0$, as expected. It is evident that the phase lag, $\delta=\alpha\pi/2$, between the stress and the strain is frequency independent; instead, it depends on the material's physical properties. The lag is less for a material with a large value of $E\theta$, which is also independently verifiable. As $\theta$ increases, the retardation time constant, $\tau=\xi/\theta$, decreases.

The study presented in this Letter agrees with the possibility of a simple process described by general principles
\cite{Miguel2002,Sethna2001}. We have shown that linking the fractional dashpot with an analogous mechanical model gave explicit physical interpretations to the Andrade law and the Omori law, which also agree with the established observations.  Surprisingly, the property of viscosity is not required to describe the two laws. Instead, the property of time-varying jerkity, along with the property of elasticity, is invoked to derive the laws. The findings may boost confidence in those results in which the constant-$Q$ model and the power-law velocity dispersion have been linked with the fractality of the medium \cite{Kjartansson1979,Lambert2015,Baan2002,Parker2018,Kawada2006}. Further, this opens the possibility of connecting the rheological properties of a medium with its geometrical heterogeneity.

In contrast to the previously proposed nonlinear theories, the methodology adopted in this Letter is of linear fractional derivatives. This is a potential paradigm shift in how the memory exhibits of a physical phenomenon can be modeled as a nonlinear system using integer-order derivatives and a linear system using fractional derivatives. However, the framework of linear fractional derivatives rules out the possibility of any chaotic behavior. A possible critique could be that the model used in obtaining the laws is phenomenological, and they lack a connection with the microdynamics of the processes. On a similar note, we cannot ignore that the power-law creep is well established
in soft metals and amorphous materials, independent of their material-specific microstructures. Besides, all macroscopic properties of a material may not always be deductible, even if a complete description of the microscopic interactions is available \cite{Cubitt2015}. Using a lumped-parameter model to describe the power laws is similar to applying mean field theories to study complex problems. An example is the utility of the Navier-Stokes equation to study fluid dynamics; the equation is largely independent of the complexities that most fluids exhibit at microscopic scales.

The author would like to thank the reviewers for their insightful comments and efforts toward improving the quality of the manuscript.

\bibliographystyle{unsrt}

\end{document}